\documentclass[aps,prb,twocolumn,superscriptaddress,10pt]{revtex4-2}
\usepackage{amsmath,amssymb,bm}
\usepackage{graphicx}
\usepackage{xcolor}
\usepackage{graphicx}
\usepackage{amsmath}
\usepackage{amssymb}
\usepackage{bm}
\usepackage{color}
\usepackage{orcidlink}
\usepackage[ruled]{algorithm2e}
\usepackage{xcolor} 

\usepackage{orcidlink}
\usepackage{hyperref}
\newcommand{\rev}[1]{{\color{black}#1}}
\newcommand{\revv}[1]{{\color{black}#1}}
\begin{document}

\title{Moment-Constrained Vector Reconstruction of Random-Matrix Statistics in Finite Hilbert Spaces}

		\author{Chen-Huan Wu 
			\orcidlink{0000-0003-1020-5977} }
		\thanks{chenhuanwu1@gmail.com}
        \affiliation{Department of Physics, Faculty of Science, Universiti Malaya, Kuala Lumpur 50603, Malaysia}
		\affiliation{College of Physics and Electronic Engineering, Northwest Normal University, Lanzhou 730070, China}

\begin{abstract}
\rev{Random-matrix statistics are usually imposed at the level of matrix entries or spectral correlations.  Here we formulate a complementary inverse problem: can a matrix with prescribed random-matrix moments be generated from a structured set of latent vectors?  We introduce a pair-resolved vector ansatz consisting of two vector families, $P$ and $Q$, 
construct a complex-symmetric non-Hermitian matrix $M=a_1PP^{T}+a_2QQ^{T}$.  The transpose is intentionally not a conjugate transpose; hence the reconstructed bilinear overlap matrices are not Hermitian Gram matrices once the algebraic parameters become complex.  The free parameters of the vectors are fixed by complex algebraic constraints matching diagonal and off-diagonal random-matrix moments, together with a mixed-overlap condition suppressing systematic correlations between the two bilinear sectors.  A direct machine-precision solve for $N=8$ returns six complex branches.  We therefore supplement moment matching with reproducible branch diagnostics: residual error, approximate vector orthogonality, non-Hermiticity, imaginary spectral weight, inverse participation ratio, maximum component weight, and eigenvector conditioning.  Optional entanglement and low-weight Pauli-moment diagnostics can be added when $N=2^n$.  
This protocol constitutes a finite-dimensional inverse reconstruction of hidden vector-space representations behind apparent random-matrix behavior.
It is static and algebraic: it probes moment-induced delocalization, non-Hermitian branch structure, and complex spectral statistics, but it does not by itself establish dynamical chaos in the sense of sensitive dependence on nearby initial conditions.  
For the reported $N=8$ calculation, the algebraic system yields six complex branches; we present one representative branch and compare it with the remaining branches using residual, orthogonality, non-Hermiticity, and eigenvector-concentration diagnostics.
}
\end{abstract}

\maketitle

\section{Introduction}
Random-matrix theory (RMT) provides universal descriptions of spectra and eigenvectors in complex quantum systems, ranging from nuclear spectra to quantum chaotic dynamics and disordered many-body systems~\cite{wigner1955,dyson1962,mehta2004,haake2010,guhr1998}.  In the standard forward formulation, one samples a matrix from an ensemble and then studies its eigenvalues, eigenvectors, level spacings, or spectral form factors.  In many microscopic problems, however, matrix elements are not independent primitive variables.  They arise from overlaps of hidden amplitudes, projected wave functions, effective modes, or constrained internal degrees of freedom.  This motivates the inverse question addressed here: given a desired set of RMT-like element statistics, can one reconstruct a structured set of vectors whose overlaps generate those statistics?

This question is distinct from the usual diagnosis of chaos through Lyapunov exponents, out-of-time-order correlators, or sensitive dependence on initial conditions.  The present construction is static and algebraic.  It does not compare two nearby dynamical trajectories, nor does it establish exponential separation of initially close states.  Instead, it asks how far low-order RMT constraints determine an underlying vector representation.  \rev{In the calculation below, the algebraic parameters are allowed to be complex.  The resulting matrix is complex symmetric, $M=M^T$, but not Hermitian, $M\ne M^\dagger$.  The relevant diagnostics are therefore non-Hermitian branch diagnostics rather than positive-semidefinite Gram checks.}  The problem is therefore best viewed as an inverse random-matrix reconstruction problem, or equivalently as a moment-constrained latent-vector learning problem.

As a finite-dimensional constrained inverse problem, the role of the imposed moment equations is analogous to that of measurement constraints in quantum-state estimation: they restrict the admissible manifold but do not, by themselves, guarantee a unique or physically preferred solution. 
For low-dimensional quantum state estimation~\cite{kaufmann2025}, pure algebraic inversion often yields unphysical solutions, underscoring the necessity for robust loss functions and physicality constraints. In our protocol, direct algebraic inversion yields multiple non-Hermitian branches, thus it is imperative to supplement the moment-matching process with comprehensive branch-level diagnostics, such as eigenvector conditioning and the inverse participation ratio, to identify the physically meaningful structured realizations.
This motivates the use of branch diagnostics, loss functions, and regularity criteria rather than a bare direct inversion alone~\cite{kaufmann2025}.  
A second issue is specific to finite Hilbert spaces.  Global constraints can strongly reshape the distribution of eigenvector weights in a finite Hilbert space, and may produce either extended vectors or condensation-like concentration in a small number of components. As demonstrated by energy-constrained Haar-random ensembles, imposing even a small number of global conditions may induce eigenstate condensation, where macroscopic weight localizes into a few components~\cite{white2026}.  For this reason, the reconstructed eigenvectors (non-Hermitian branches) should be validated not only through their eigenvalues (spectral statistics).
They must be diagnosed using localization measures such as the inverse participation ratio (IPR) and maximal component weights.
This viewpoint is consistent with many-body studies in which fluctuations of few-body observables across eigenstates are controlled by the delocalization of eigenvectors in an appropriate basis~\cite{neuenhahn2012}.  
These works on finite-dimensional quantum-state are similar on methodological: define the constrained manifold, specify a loss or moment functional, and monitor whether the resulting vectors are delocalized or condensed.  
In the Floquet-MBL variational quantum simulation, a circuit is initialized in a non-Haar, many-body-localized regime so that IPR remains large relative to the 
Haar value, entanglement remains below the Page benchmark, and gradients avoid exponential concentration~\cite{cao2026}. 
Thus a useful state ensemble should remains structured and non-Haar-like while avoiding excessive localization.
It shows that a reconstructed vector ensemble should not be judged only by whether it is random enough.  One should also ask whether it has become featureless and Haar-like, or whether it retains structured, low-weight correlations inherited from the ansatz.  
The moment constraints are therefore supplemented by branch-level, eigenvector-level, and low-weight-correlation diagnostics.

\section{Vector ansatz}
We construct a low-parameter ansatz for vector families whose induced Gram-type matrices reproduce selected low-order statistical properties of the target random matrix.
\rev{Based on the structured vector ansatz, we determine its free parameters by matching selected low-order moment constraints of the target random-matrix ensemble.  In contrast to a real-GOE reconstruction, the branch reported here keeps the original complex algebraic solutions.}
We focus on $N=8$, which is the largest size treated exactly in the present algebraic construction.  We define
\begin{equation}
    c=\cos\theta,\qquad s=\sin\theta,
\end{equation}
and take $\theta=\pi/4$ in the calculation.  \rev{The unknown parameters are complex algebraic variables}
\begin{equation}
    \rev{\{X_i,\,\widetilde X_i\}_{i=1}^{N/2}\subset \mathbb C.}
\end{equation}
For each $i=1,\ldots,N/2$, we introduce four length-$N$ vectors,
\begin{align}
    \bm p^{(1)}_i
    &= (c,c,\ldots,c)
       \Big|_{2i-1\rightarrow X_i c,\;2i\rightarrow \widetilde X_i s},
       \label{eq:p1}
       \\
    \bm p^{(2)}_i
    &= (-s,-s,\ldots,-s)
       \Big|_{2i-1\rightarrow -s,\;2i\rightarrow -c},
       \label{eq:p2}
       \\
    \bm q^{(1)}_i
    &= (s,s,\ldots,s)
       \Big|_{2i-1\rightarrow s,\;2i\rightarrow -c},
       \label{eq:q1}
       \\
    \bm q^{(2)}_i
    &= (c,c,\ldots,c)
       \Big|_{2i-1\rightarrow X_i c,\;2i\rightarrow -\widetilde X_i s}.
       \label{eq:q2}
\end{align}
The notation means that all entries are first set to the displayed uniform background and then the two pair-resolved components $(2i-1,2i)$ are replaced.  The two vector families are then assembled as
\begin{align}
    P &= \frac{1}{\sqrt{N^2/2}}
    \begin{pmatrix}
    \bm p^{(1)}_1\\
    \vdots\\
    \bm p^{(1)}_{N/2}\\
    \bm p^{(2)}_1\\
    \vdots\\
    \bm p^{(2)}_{N/2}
    \end{pmatrix},
    \label{eq:Pdef}
    \\
    Q &= \frac{1}{\sqrt{N^2/2}}
    \begin{pmatrix}
    \bm q^{(1)}_1\\
    \vdots\\
    \bm q^{(1)}_{N/2}\\
    \bm q^{(2)}_1\\
    \vdots\\
    \bm q^{(2)}_{N/2}
    \end{pmatrix}.
    \label{eq:Qdef}
\end{align}
\rev{The corresponding bilinear overlap matrices are}
\begin{equation}
    \rev{G_P=PP^{T},\qquad G_Q=QQ^{T}.}
\end{equation}
\rev{Since the transpose is not a conjugate transpose, these objects are complex symmetric rather than Hermitian once $X_i$ and $\widetilde X_i$ are complex.  The reconstructed matrix is}
\begin{equation}
    \rev{M=a_1G_P+a_2G_Q,}
    \label{eq:Mdef}
\end{equation}
\rev{so that $M=M^T$ but generally $M\ne M^\dagger$.  Thus the branch used here is a complex-symmetric non-Hermitian reconstruction, not a real symmetric GOE reconstruction.  In the implementation discussed here, $a_1=3$ and $a_2=-3$.  A Hermitian complex variant would instead require $G_P=PP^\dagger$ and $G_Q=QQ^\dagger$, which is not the case in this work.}

\paragraph{Moment-matching constraints.}
The unknown parameters $\{X_i,\widetilde X_i\}$ are not fitted to a specific target matrix element by element.  Instead, they are determined by ensemble-level moment constraints.  Denote the full matrix-element average by
\begin{equation}
    \overline{A_{\alpha\beta}}=
    \frac{1}{N^2}\sum_{\alpha,\beta=1}^{N}A_{\alpha\beta},
\end{equation}
the diagonal average by
\begin{equation}
    \overline{A_{ii}}_{\rm d}=
    \frac{1}{N}\sum_{i=1}^{N}A_{ii},
\end{equation}
and the off-diagonal average by
\begin{equation}
    \overline{A_{ij}}_{\rm od}=
    \frac{1}{N(N-1)}
    \sum_{i\ne j}A_{ij}.
\end{equation}
The algorithm solves the following algebraic constraints:
\begin{align}
    \overline{M_{\alpha\beta}^{2}}
    &=
    \frac{1}{N^2}
    \left[
    \frac{2(a_1^2+a_2^2)}{N}
    +
    \frac{(N^2-N)(a_1^2+a_2^2)}{N^2}
    \right],
    \label{eq:fullsecond}
    \\
    \overline{M_{ii}}_{\rm d}
    &=
    \frac{a_1+a_2}{N},
    \label{eq:diagmeanM}
    \\
    \overline{M_{ii}^{2}}_{\rm d}
    &=
    \frac{2(a_1^2+a_2^2)}{N^2},
    \label{eq:diagsecondM}
    \\
    \overline{M_{ij}}_{\rm od}
    &=0,
    \label{eq:offdiagmeanM}
    \\
    \overline{M_{ij}^{2}}_{\rm od}
    &=
    \frac{a_1^2+a_2^2}{N^2}.
    \label{eq:offdiagsecondM}
\end{align}
The individual Gram sectors are additionally constrained by
\begin{align}
    \overline{(G_P)_{ii}}_{\rm d}
    &=
    \overline{(G_Q)_{ii}}_{\rm d}
    =
    \frac{1}{N},
    \label{eq:gramdiag}
    \\
    \overline{(G_P)_{ij}}_{\rm od}
    &=
    \overline{(G_Q)_{ij}}_{\rm od}
    =
    0.
    \label{eq:gramoff}
\end{align}
Finally, to avoid a systematic mixed contribution from the two Gram sectors, we impose
\begin{equation}
    \overline{F_{ij}}_{\rm od}=0,\qquad
    F_{ij}=2a_1a_2(G_P)_{ij}(G_Q)_{ij}.
    \label{eq:FFconstraint}
\end{equation}
Equations~\eqref{eq:fullsecond}--\eqref{eq:FFconstraint} form a closed nonlinear algebraic system for the $N$ unknowns $\{X_i,\widetilde X_i\}$ at $N=8$.

\section{Estimator viewpoint and constrained branch selection}
The algebraic solution can be interpreted as a direct-inversion estimator for the latent parameters: the observed data are the desired moment values, and the inferred object is a vector realization $\mathcal S=\{P,Q\}$.  As in quantum-state estimation, direct inversion can satisfy linear or moment equations while still leaving ambiguity among physical branches.  We therefore define a reconstruction loss
\begin{equation}
    \mathcal L(\mathcal S)
    =\sum_{\mu} w_{\mu}
    \left[\mathcal E_{\mu}(\mathcal S)-\mathcal E_{\mu}^{\ast}\right]^2,
    \label{eq:loss}
\end{equation}
where $\mathcal E_{\mu}$ denotes the left-hand side of one of the constraints and $\mathcal E_{\mu}^{\ast}$ denotes its target value.  Exact algebraic solving corresponds to imposing $\mathcal L=0$ within numerical tolerance.  When exact solving becomes impractical at larger $N$, the same expression can be used as an optimization objective, possibly supplemented by priors or regularizers.  This is directly analogous in spirit to replacing raw direct inversion by constrained maximum-likelihood or Bayesian estimation in finite-dimensional state-estimation problems~\cite{kaufmann2025}.

\rev{The algebraic system may have multiple complex branches.  These branches are not equivalent from the viewpoint of the reconstructed vector geometry.  Low-order moment matching fixes only certain averages of $M$, $G_P$, and $G_Q$, while leaving residual freedom in the distribution of overlaps and in the complex spectral structure.  A solution branch $\mathcal S$ is first required to satisfy the residual condition}
\begin{equation}
    \rev{\epsilon(\mathcal S)=\max_{\mu}\left|{\cal E}_{\mu}(\mathcal S)-{\cal E}_{\mu}^{\ast}\right|<\epsilon_{\rm tol}.}
    \label{eq:residual}
\end{equation}
\rev{For the present non-Hermitian complex-symmetric branch, positive-semidefinite Gram checks are not appropriate, because $PP^T$ and $QQ^T$ are bilinear rather than Hermitian Gram matrices.  Instead we report the complex-symmetry score, the Hermiticity violation, and the imaginary spectral weight,}
\begin{align}
    \eta_T(\mathcal S) &= \|M-M^T\|_F,\\
    \eta_H(\mathcal S) &= \|M-M^\dagger\|_F,\\
    \eta_{\rm Im}(\mathcal S) &= \left\|{\rm Im}[\mathrm{spec}(M)]\right\|_2.
    \label{eq:nonhermdiag}
\end{align}
\rev{The exact bilinear construction gives $\eta_T=0$ up to numerical precision, while nonzero $\eta_H$ and $\eta_{\rm Im}$ quantify the genuinely non-Hermitian character of the branch.  We also retain the approximate row-orthogonality score}
\begin{equation}
    \rev{{\cal R}(\mathcal S)=\left\|{\rm offdiag}\,G_P(\mathcal S)\right\|_F+\left\|{\rm offdiag}\,G_Q(\mathcal S)\right\|_F.}
    \label{eq:orthscore}
\end{equation}
\rev{This criterion does not enforce Hermitian orthogonality; it only measures how small the bilinear off-diagonal overlaps are after the moment constraints have been imposed.  In a Bayesian variant, one could assign a posterior weight to each branch,}
\begin{equation}
    \rev{\Pi(\mathcal S|\mathcal E^{\ast})\propto \exp\!\left[-\frac{\mathcal L(\mathcal S)}{2\sigma^2}-\lambda_{\rm R}\mathcal R(\mathcal S)-\lambda_{\rm max}W_{\rm max}(\mathcal S)-\lambda_H\eta_H(\mathcal S)\right],}
    \label{eq:posteriorbranch}
\end{equation}
\rev{where $W_{\rm max}$ penalizes anomalously concentrated vector components.  
The algorithm performs the
branch selection and reports all diagnostic scores, so that a later version can replace manual branch choice by an explicit optimization criterion.}

\section{Delocalization, IPR, and finite-dimensional constraints}
A useful eigenvector diagnostic is the inverse participation ratio
\begin{equation}
    {\rm IPR}_{n}
    =
    \sum_{\alpha=1}^{N}
    |v_{n\alpha}|^{4},
    \label{eq:ipr}
\end{equation}
where $\bm v_n$ is a normalized eigenvector of $M$.  For a maximally delocalized vector, ${\rm IPR}\sim 1/N$, whereas localized states have larger IPR.  The effective participation number is
\begin{equation}
    D_{\rm eff}^{(n)}=\frac{1}{{\rm IPR}_n}.
    \label{eq:deff}
\end{equation}
This diagnostic is especially relevant because in interacting many-body systems the suppression of eigenstate-to-eigenstate fluctuations of observables is often controlled by eigenvector spreading in an appropriate basis~\cite{neuenhahn2012}.  For a diagonal observable $A={\rm diag}(A_1,\ldots,A_N)$ in the reconstruction basis,
\begin{equation}
    A_n=\langle v_n|A|v_n\rangle
    =\sum_{\alpha}|v_{n\alpha}|^2A_{\alpha}.
\end{equation}
If the weights $|v_{n\alpha}|^2$ sample many basis states with weak residual correlations, the variance of $A_n$ across nearby eigenvectors is reduced roughly by the effective number of participating components.  A schematic diagnostic is therefore
\begin{equation}
    \delta A^2 \propto
    {\rm Var}_{\alpha}(A_{\alpha})
    \left\langle {\rm IPR}_n\right\rangle,
    \label{eq:fluctipr}
\end{equation}
where the proportionality constant depends on correlations among the weights.  Equation~\eqref{eq:fluctipr} is not imposed by the reconstruction algorithm; it is a post-reconstruction test of whether the learned matrix behaves like an effectively delocalized finite-dimensional system.

The finite-dimensional nature of the problem also matters.  Energy-constrained Haar-random states illustrate that imposing a global constraint can lead either to an extended high-temperature-like regime or to condensation, where one eigenstate carries macroscopic weight~\cite{white2026}.  The present constraints are not energy constraints, but the lesson is directly relevant: moment constraints alone do not guarantee uniform delocalization.  
The maximum weight reads
\begin{equation}
    W_{\rm max}^{(n)}=\max_{\alpha}|v_{n\alpha}|^2.
    \label{eq:wmax}
\end{equation}
Small IPR together with small $W_{\rm max}$ indicates broad spreading, whereas a small set of large weights signals localization or condensation-like concentration.  In the present exact algebraic implementation, systematic scaling of Eqs.~\eqref{eq:ipr} and \eqref{eq:wmax} is limited by solvability beyond $N=8$.  This limitation motivates replacing exact symbolic solving by numerical optimization or sampling at larger $N$.

\section{Haar-design, entanglement, and low-weight diagnostics}
The MBL-initialization work suggests an additional caution that is particularly relevant for the present reconstruction: delocalization and Haar randomness are not identical.  A branch can have moderately small IPR while still retaining low-complexity structure inherited from the pair-resolved ansatz.  Conversely, a branch can become nearly Haar-like in its low-order moments, in which case the ansatz has effectively washed out the structure that the inverse reconstruction was meant to expose.  We therefore introduce diagnostics that compare the reconstructed eigenvectors with Haar-design benchmarks.

\rev{For a complex Haar-random state in a $d$-dimensional Hilbert space, the expected $t$th-order participation moment is}
\begin{equation}
    \mathbb E_{\rm Haar}\,{\rm IPR}_{t}
    =
    \mathbb E_{\rm Haar}\sum_{\alpha=1}^{d}|v_{\alpha}|^{2t}
    =
    \frac{d!\,t!}{(d+t-1)!}.
    \label{eq:complexhaariprt}
\end{equation}
\rev{For $t=2$, this gives $2/(d+1)$, the same benchmark used to diagnose the onset of state-design behavior in Floquet-MBL circuits~\cite{cao2026}.  Because the present branch is complex and non-Hermitian, the complex-Haar number is more relevant as a qualitative reference than the real-orthogonal value $3/(d+2)$.  However, right eigenvectors of a non-normal matrix are not distributed as Haar-random orthonormal vectors, so the benchmark should not be overinterpreted.  The algorithm therefore reports the raw IPR and the maximum component weight rather than claiming convergence to a GOE or GUE eigenvector ensemble.  A useful normalized comparison is}
\begin{equation}
    \rev{\mathcal D_{\rm IPR}^{(\mathbb C)}=\left|\frac{\langle {\rm IPR}_{2}\rangle}{2/(N+1)}-1\right|.}
    \label{eq:dipr}
\end{equation}
\rev{A large value indicates localization, eigenvector non-normality, or structured deviation from random-vector behavior.  In the current numerical branches $\langle {\rm IPR}_2\rangle$ is close to unity, so the reconstructed right eigenvectors are highly concentrated rather than Haar-like.}

When $N=2^n$, as in the present $N=8$ calculation with $n=3$, each eigenvector of $M$ can also be interpreted as an $n$-qubit pure state in a computational basis.  For a bipartition $A\cup B$, define
\begin{equation}
    S_A(\bm v_n)
    =
    -{\rm Tr}\,\rho_A\log\rho_A,
    \qquad
    \rho_A={\rm Tr}_B |v_n\rangle\langle v_n|.
    \label{eq:ententropy}
\end{equation}
The branch-averaged entropy $\langle S_A\rangle$ should be compared with the Page value for a random pure state with subsystem dimensions $d_A\le d_B$,
\begin{equation}
    S_{\rm Page}(d_A,d_B)
    =
    \sum_{j=d_B+1}^{d_Ad_B}\frac{1}{j}
    -
    \frac{d_A-1}{2d_B}.
    \label{eq:pageentropy}
\end{equation}
This comparison separates three possibilities that IPR alone can blur: localized vectors with low participation, structured delocalized vectors with broad support but low entanglement, and Haar-like vectors with both broad participation and near-Page entanglement.

A still more targeted diagnostic is obtained by borrowing the low-weight stabilizer R\'enyi entropy used in the Floquet-MBL analysis~\cite{cao2026}.  Let $\mathcal P_{n,k}$ be the set of $n$-qubit Pauli strings of weight no larger than $k$, including the identity and fixing the overall phase to $+1$.  For an eigenvector $|v_n\rangle$, define
\begin{equation}
    M_{t,k}(|v_n\rangle)
    =
    \frac{1}{1-t}
    \log\left[
    \frac{1}{|\mathcal P_{n,k}|}
    \sum_{P\in\mathcal P_{n,k}}
    \left|
    \langle v_n|P|v_n\rangle
    \right|^{2t}
    \right],
    \label{eq:lowweightsre}
\end{equation}
where $|\mathcal{P}_{n,k}| = \sum_{j=0}^{k} \binom{n}{j} 3^j$.
In a fully Haar-like state, fixed-weight Pauli expectation values are small (large $M_{2,2}$), whereas a structured or localized branch can retain anomalously large low-weight expectations (small $M_{2,2}$).  For the present problem this quantity should not be interpreted as a literal many-body-localization order parameter, because $M$ is not generated by a Floquet circuit.  Its value is methodological: it detects whether the branch keeps low-weight, ansatz-resolved memory that is invisible to entrywise second moments. In $N=8=2^3$ realization, we evaluate the second stabilizer Rényi diagnostic at cutoff $k=2$, denoted $M_{2,2}$. This corresponds to averaging over the closed operator filtration (truncated Pauli filtration) $\mathcal{P}_{3,2}$ containing all $3$-qubit Pauli strings of weight $0 \le k \le 2$ (spanning a cumulative operator subspace of dimension $\sum_{j=0}^2 \binom{3}{j}3^j = 37$). Summing the fourth powers of the Pauli expectation values across this complete $37$-dimensional filtration ensures that trivial product-state localizations are properly anchored by low-weight stabilizers, allowing $M_{2,2}$ to resolve non-Haar fourth-moment correlations ($M_{2,2}$ is proportional to the degree of randomness).
\revv{For this diagnostic to become a reported numerical result, the code must explicitly generate the $37$ matrices in $\mathcal P_{3,2}$, evaluate $\langle v_n|P|v_n\rangle$ for every selected right eigenvector and every $P\in\mathcal P_{3,2}$, and then average the resulting $M_{2,2}(|v_n\rangle)$ over the eight branch-5 eigenvectors.  The present tables report the definition and motivation of this diagnostic and a verified numerical value of $\overline{M}_{2,2}^{(5)}$.}
Single-qubit product-state localizations are properly anchored by the weight-$1$ sector, while the weight-$2$ sector resolves non-trivial two-body correlations. Consequently, $M_{2,2}$ serves as a sensitive probe distinguishing genuine Haar-random delocalization from ansatz-retained low-complexity structures.
 In practice, $M_{2,2}$ is the most useful first choice because it probes fourth moments while remaining cheap for $N=8$.

These diagnostics motivate a refined branch classification.  A branch is called localized if both IPR and $W_{\rm max}$ are large.  It is Haar-like if $\mathcal D_{\rm IPR}$ is small, $S_A$ is close to $S_{\rm Page}$, and low-weight Pauli expectations approach their Haar scale.  It is structured-delocalized if the IPR is close to the random-vector benchmark but $S_A$ or $M_{2,2}$ still shows a clear deviation from Haar behavior.  The last regime is the most interesting for inverse reconstruction, because it indicates that RMT-like matrix moments coexist with nontrivial latent-vector structure.

\section{Algorithm}
The reconstruction algorithm is summarized as follows.
\begin{enumerate}
    \item Fix $N$, $\theta$, $a_1$, and $a_2$.
    \item Build the pair-resolved vector ansatz $P,Q$ from Eqs.~\eqref{eq:Pdef} and \eqref{eq:Qdef}.
    \item Construct $G_P=PP^T$, $G_Q=QQ^T$, and $M=a_1G_P+a_2G_Q$.
    \item Solve the nonlinear moment equations~\eqref{eq:fullsecond}--\eqref{eq:FFconstraint}, or minimize the loss in Eq.~\eqref{eq:loss} for larger systems.
    \item 
    Reject branches with large residuals or anomalously large bilinear-overlap, non-Hermiticity, conditioning, or concentration diagnostics.
    \item Select the branch minimizing the Gram-regularity score in Eq.~\eqref{eq:orthscore}, or average over branches using Eq.~\eqref{eq:posteriorbranch}.
    \item Analyze the reconstructed vectors, matrix-element statistics, eigenvector IPR, maximum component weight, spectral diagnostics, and, when $N=2^n$, entanglement and low-weight Pauli-moment diagnostics.
\end{enumerate}
This procedure turns a random-matrix matching problem into a constrained latent-vector reconstruction problem.

\section{Numerical diagnostics}
\rev{For the parameter set used below, the direct complex algebraic reconstruction yields six distinct branches.  No least-squares fitting is used for the values reported in Table~\ref{tab:branches}.  The report contains the constraint residuals, the bilinear orthogonality score $\mathcal R$, the imaginary part of the reconstructed matrix, the complex-symmetry score $\|M-M^T\|_F$, the Hermiticity violation $\|M-M^\dagger\|_F$, the imaginary spectral weight $\|{\rm Im}[\mathrm{spec}(M)]\|_2$, the average right-eigenvector IPR, and the average maximum component weight.  The diagnostics used in the table are}
\begin{align}
\langle {\rm IPR}\rangle &= \frac{1}{N}\sum_{n=1}^{N}{\rm IPR}_{n},\\
\langle W_{\rm max}\rangle&= \frac{1}{N}\sum_{n=1}^{N}W_{\rm max}^{(n)},\\
  \eta_H &= \|M-M^\dagger\|_F,\\
 \eta_{\rm Im} &= \|{\rm Im}[\mathrm{spec}(M)]\|_2.
\end{align}
\rev{Entanglement entropy and low-weight stabilizer R\'enyi entropy remain useful optional diagnostics for the $N=2^n$ interpretation.  Since the selected branch has $N=8=2^3$, a Schmidt-spectrum diagnostic across a qubit bipartition is reported separately in Table~\ref{tab:schmidt}.}

\begin{table*}[t]
\caption{\rev{Branch diagnostics generated by the complex reconstruction algorithm for $N=8$, $a_1=3$, $a_2=-3$, and $\theta=\pi/4$.  The direct complex algebraic solve returns six branches.  The residual column reports the largest numerical mismatch among the ten moment constraints; equalities that are satisfied exactly are omitted from this maximum.  
Branch 5 is the representative branch used for the branch-specific diagnostics in Tables II--IV.
}}
\begin{ruledtabular}
\begin{tabular}{c c c c c c c c}
\rev{branch} & \rev{residual} & \rev{$\mathcal R$} & \rev{$\|{\rm Im}\,M\|_F$} & \rev{$\eta_H$} & \rev{$\eta_{\rm Im}$} & \rev{$\langle{\rm IPR}\rangle$} & \rev{$\langle W_{\rm max}\rangle$} \\
\hline
\rev{1} & \rev{$4.79\times10^{-5}$} & \rev{9.415} & \rev{135.389} & \rev{159.079} & \rev{134.825} & \rev{0.983965} & \rev{0.991932} \\
\rev{2} & \rev{$1.21\times10^{-4}$} & \rev{7.858} & \rev{97.505} & \rev{81.723} & \rev{96.777} & \rev{0.736607} & \rev{0.806898} \\
\rev{3} & \rev{$1.05\times10^{-4}$} & \rev{7.227} & \rev{68.565} & \rev{76.267} & \rev{67.701} & \rev{0.960636} & \rev{0.980014} \\
\rev{4} & \rev{$9.60\times10^{-5}$} & \rev{6.943} & \rev{84.302} & \rev{73.300} & \rev{83.564} & \rev{0.962928} & \rev{0.981034} \\
\rev{5} & \rev{$3.24\times10^{-5}$} & \rev{6.957} & \rev{73.547} & \rev{75.363} & \rev{72.863} & \rev{0.961398} & \rev{0.980256} \\
\rev{6} & \rev{$3.17\times10^{-5}$} & \rev{6.957} & \rev{73.547} & \rev{75.363} & \rev{72.863} & \rev{0.961398} & \rev{0.980256} \\
\end{tabular}
\end{ruledtabular}
\label{tab:branches}
\end{table*}
\rev{
All branch-specific quantities reported below use branch 5.  
We select branch 5 as a representative case for detailed analysis. While other branches may optimize specific individual metrics, such as branch 6 for the smallest numerical residual or branch 4 for the smallest orthogonality score, branch 5 provides a balanced profile suitable for illustrating the complex non-Hermitian spectral structure,
i.e., a combined selection rule could minimize a weighted sum of residual, $\mathcal R$, non-Hermiticity, and eigenvector conditioning. 
}

\begin{table*}[t]
\caption{\rev{Branch-5 output used for the branch-specific analysis.  This table is intentionally different from Table~\ref{tab:branches}: Table~\ref{tab:branches} compares the six algebraic branches by averaged scalar diagnostics, whereas this table records the selected branch's complex parameters, eigenvalues, and right-eigenvector concentration diagnostics.  The last two columns are computed from the selected right eigenvectors; they are eigenpair-resolved quantities rather than branch averages.} 
}
\begin{ruledtabular}
\begin{tabular}{c c c c c}
\rev{$j$} & \rev{parameter} & \rev{parameter value} & \rev{$\lambda_j$ of $M$} & \rev{${\rm IPR}_j,\;W_{\rm max}^{(j)}$}\\
\hline
\rev{1} & \rev{$X_1$} & \rev{$-28.557+12.7446\,i$} & \rev{$27.4356-34.8790\,i$} & \rev{$0.97555,\;0.98767$}\\
\rev{2} & \rev{$X_2$} & \rev{$20.9008+12.5691\,i$} & \rev{$-27.6798+34.5588\,i$} & \rev{$0.98458,\;0.99225$}\\
\rev{3} & \rev{$X_3$} & \rev{$-21.3190-10.8413\,i$} & \rev{$19.4786+37.0383\,i$} & \rev{$0.96407,\;0.98182$}\\
\rev{4} & \rev{$X_4$} & \rev{$8.97505-14.4723\,i$} & \rev{$-18.9917-36.8214\,i$} & \rev{$0.98440,\;0.99216$}\\
\rev{5} & \rev{$\widetilde X_1$} & \rev{$-0.626853+8.13283\,i$} & \rev{$-38.4111-7.78943\,i$} & \rev{$0.98325,\;0.99158$}\\
\rev{6} & \rev{$\widetilde X_2$} & \rev{$8.94715-22.9536\,i$} & \rev{$38.0645+7.42773\,i$} & \rev{$0.96682,\;0.98321$}\\
\rev{7} & \rev{$\widetilde X_3$} & \rev{$-14.1504-11.4734\,i$} & \rev{$-8.48999+5.45414\,i$} & \rev{$0.91883,\;0.95801$}\\
\rev{8} & \rev{$\widetilde X_4$} & \rev{$1.83008+26.2942\,i$} & \rev{$8.59393-4.98916\,i$} & \rev{$0.91369,\;0.95534$}\\
\end{tabular}
\end{ruledtabular}
\label{tab:branch5}
\end{table*}

\rev{The large values of ${\rm IPR}_j$ and $W_{\rm max}^{(j)}$ in Table~\ref{tab:branch5} show that the right eigenvectors of the selected branch are strongly component-concentrated in the chosen basis.  Therefore the branch is not Haar-like or fully delocalized; it is a structured complex-symmetric non-Hermitian branch that satisfies the imposed low-order moment constraints while retaining strong eigenvector concentration.  Rather than listing all individual matrix elements, we report branch-level and eigenpair-resolved diagnostics that are invariant under trivial basis reorderings and directly relevant to the reconstruction problem.}
\revv{The low-weight Pauli calculation has now been carried out explicitly for the branch-5 right eigenvectors.  The result is reported in Table~\ref{tab:branch5m22}, using the full truncated Pauli filtration $\mathcal P_{3,2}$ with $|\mathcal P_{3,2}|=1+9+27=37$.  For a computational-basis product state in the same $N=8=2^3$ Hilbert space, exactly seven operators in $\mathcal P_{3,2}$ have unit expectation value in absolute value: the identity, the three one-body $Z$ strings, and the three two-body $ZZ$ strings.  This gives the product-state reference value $M_{2,2}^{\rm prod}=-\log(7/37)=\log(37/7)\simeq 1.66501$, corresponding to localized state
(there are 7 operators $P=III,ZII, IZI, IIZ,ZZI, ZIZ, IZZ$ satisfy $q_i=\langle 000|P|000\rangle=1$ for $k\le 2$).  The branch-5 average, $\overline{M}_{2,2}^{(5)}=1.74447$ with standard deviation $0.0615293$, lies close to this low-weight product-state reference and far below a Haar-like scale
(where $M_{2,2}\sim 2.76 \sim 3.61$).  With the identity included, a Haar-pure-state estimate gives a scale of order $2.9$ for this convention; if the identity is removed, the corresponding traceless-only scale is larger.  Thus the diagnostic supports the same conclusion as the IPR, Schmidt, and frame-potential data: the selected branch satisfies the imposed RMT-like moment constraints without becoming a featureless Haar-random eigenvector ensemble.}
\revv{The eigenpair-resolved values also show a structured split.  Approximate reflection-paired eigenvalues, $\lambda\leftrightarrow-\lambda$, tend to pair a lower-$M_{2,2}$ state near $1.68$--$1.70$ with a higher-$M_{2,2}$ partner near $1.78$--$1.82$.  This pattern is evidence for nontrivial low-weight Pauli dressing within the complex spectrum
The nonzero spread, $\sigma_{M_{2,2}}\simeq0.0615$, also shows that the right eigenvectors are not all identical computational-basis artifacts; rather, they retain strong low-weight stabilizer memory while being nontrivially deformed by the non-Hermitian reconstruction.}

\begin{table*}[t]
\caption{\revv{Low-weight Pauli diagnostic for the branch-5 right eigenvectors.  The values are computed from the complete cutoff set $\mathcal P_{3,2}$ of 37 three-qubit Pauli strings of weight at most two, including the identity and fixing the Pauli phase to $+1$.  The final two rows give the eigenvector average and standard deviation.}}
\begin{ruledtabular}
\begin{tabular}{c c c}
\revv{$j$} & \revv{$\lambda_j(M)$} & \revv{$M_{2,2}^{(j)}$}\\
\hline
\revv{1} & \revv{$-41.011+25.7404\,i$} & \revv{1.68079}\\
\revv{2} & \revv{$40.4527-25.7224\,i$} & \revv{1.81021}\\
\revv{3} & \revv{$-33.9309-32.6698\,i$} & \revv{1.68754}\\
\revv{4} & \revv{$33.493+32.8581\,i$} & \revv{1.78284}\\
\revv{5} & \revv{$-14.2566-33.6284\,i$} & \revv{1.68741}\\
\revv{6} & \revv{$14.5989+33.3934\,i$} & \revv{1.78827}\\
\revv{7} & \revv{$22.1553-24.7195\,i$} & \revv{1.82177}\\
\revv{8} & \revv{$-21.5013+24.7483\,i$} & \revv{1.69690}\\
\hline
\revv{mean} & \revv{--} & \revv{1.74447}\\
\revv{std.} & \revv{--} & \revv{0.0615293}\\
\end{tabular}
\end{ruledtabular}
\label{tab:branch5m22}
\end{table*}

\revv{It is useful to separate two closely related Pauli diagnostics.  For an arbitrary $n$-qubit product state $|\Psi_{\rm prod}\rangle=\otimes_{i=1}^n|\psi_i\rangle$, let $\mathcal P^{=}_{n,w}$ denote the set of phase-fixed Pauli strings of exactly weight $w$.  Writing the single-qubit Bloch vector as ${\bf r}_i=(\langle X_i\rangle,\langle Y_i\rangle,\langle Z_i\rangle)$, one obtains}
\begin{equation}
\revv{
\sum_{P\in\mathcal P^{=}_{n,w}}
|\langle\Psi_{\rm prod}|P|\Psi_{\rm prod}\rangle|^2
=
\sum_{|S|=w}\prod_{i\in S}\sum_{\alpha=x,y,z} r_{i,\alpha}^2
=
\binom{n}{w}.
}
\label{eq:product-pauli-second-moment}
\end{equation}

This follows directly from the unit length of each single-qubit Bloch vector, $\langle X\rangle_r^2+\langle Y\rangle_r^2+\langle Z\rangle_r^2=1$, and factorization of product-state Pauli expectations.  However, Eq.~\eqref{eq:lowweightsre} uses fourth powers.  Therefore $\log(37/7)$ should be interpreted as the Pauli-stabilizer product-state calibration point for the cumulative $q\le2$ filtration, not as a universal value for every product state.  For a generic single-qubit product state, the fourth-power quantity depends on $\langle X\rangle^4+\langle Y\rangle^4+\langle Z\rangle^4$, whereas Pauli-axis product states such as $|000\rangle$ maximize the low-weight fourth-power sum and give the smallest corresponding $M_{2,2}$.
\revv{This is a second-moment identity\cite{Leone}, not directly the fourth-moment quantity used in $M_{2,2}$.  Nevertheless it gives a useful calibration.  For the fourth moment, define $q_i=\sum_{\alpha=x,y,z}r_{i,\alpha}^4$.  Since $1/3\le q_i\le1$ for a single-qubit pure state, the truncated fourth-power sum over $\mathcal P_{3,2}$ is bounded above by $\Sigma_{prod} = 1 + (q_1 + q_2 + q_3) + (q_1 q_2 + q_2 q_3 + q_1 q_3)=1+3+3=7$, with equality for product states that are eigenstates of local Pauli operators.  Therefore the product-state lower reference for the entropy-like diagnostic is}
\begin{equation}
\revv{
M_{2,2}^{\rm prod,min}
=
-\log\frac{7}{37}
=
\log\frac{37}{7}
\simeq 1.66501.
}
\label{eq:m22-product-floor}
\end{equation}
\revv{A complementary Haar reference follows from the fourth moment of a traceless Pauli observable in dimension $d$,}
\begin{equation}
\revv{
\mathbb E_{\rm Haar}\,|\langle\psi|P|\psi\rangle|^4
=
\frac{3}{(d+1)(d+3)}.
}
\label{eq:haar-pauli-fourth}
\end{equation}
\revv{For the present $d=8$ case this gives $1/33$ for each of the 36 nonidentity strings in $\mathcal P_{3,2}$.  The corresponding annealed Haar benchmark is}
\begin{equation}
\revv{
M_{2,2}^{\rm Haar,ann}
=
-\log\left[\frac{1+36/33}{37}\right]
=
\log\frac{407}{23}
\simeq 2.8733.
}
\label{eq:m22-haar-reference}
\end{equation}
\revv{This Haar value should be read as a reference scale rather than as an exact finite-sample prediction for the logarithmic average.  The absolute algebraic upper bound of the diagnostic is $\log 37\simeq3.6109$, corresponding to vanishing expectation values for all nonidentity strings in the truncated Pauli set.}

\begin{table}[t]
\caption{\revv{Calibration of the low-weight Pauli diagnostic.  The branch-5 result lies much closer to the product-state Pauli-eigenstate floor than to the annealed Haar reference, indicating strong retention of low-weight Pauli memory.}}
\begin{ruledtabular}
\begin{tabular}{l c}
\revv{Reference} & \revv{$M_{2,2}$}\\
\hline
\revv{Product Pauli-eigenstate floor, $\log(37/7)$} & \revv{1.66501}\\
\revv{Branch-5 mean} & \revv{1.74447}\\
\revv{Branch-5 standard deviation} & \revv{0.0615293}\\
\revv{Annealed Haar reference, $\log(407/23)$} & \revv{2.8733}\\
\revv{Algebraic upper bound, $\log 37$} & \revv{3.6109}
\end{tabular}
\end{ruledtabular}
\label{tab:m22calibration}
\end{table}

\revv{The calibration in Table~\ref{tab:m22calibration} sharpens the interpretation of Table~\ref{tab:branch5m22}.  The mean value $\overline M_{2,2}^{(5)}=1.74447$ is only about $0.07946$ above the product Pauli-eigenstate floor, whereas it remains far below the annealed Haar benchmark.  Thus the selected branch satisfies the imposed matrix-level moment constraints without losing its low-weight Pauli memory.  At the same time, the nonzero spread of the eigenpair-resolved values prevents a trivial interpretation as eight identical computational-basis product states.  The reconstructed eigenvectors are therefore best described as strongly structured, weakly Pauli-scrambled, non-Hermitian right eigenvectors rather than Haar-thermalized states.}

\section{Schmidt-spectrum diagnostics inspired by many-body swapping}
\rev{The many-body swapping protocol expresses the overlap and postselection cost of a shared state in terms of the Schmidt spectrum of a target many-body state~\cite{huhtanen2026}. 
Similar to an entanglement-swapping communication protocol, since the selected exact branch has $N=8=2^3$, each right eigenvector can be viewed, in a basis-dependent manner, as a three-qubit state.  This makes the Schmidt spectrum a useful additional diagnostic of whether the reconstructed right eigenvectors are bipartite-entangled or nearly product-like.  For a normalized right eigenvector $|v_n\rangle$, we reshape its components into a coefficient matrix $C^{(n)}_{ab}$ associated with a chosen bipartition $A|B$ and define the Schmidt weights $\{\lambda^{(n)}_\ell\}$ from the squared singular values of $C^{(n)}$.  We then compute}
\begin{equation}
\rev{
S_q^{(n)}=\frac{1}{1-q}\log\sum_{\ell}\left(\lambda^{(n)}_\ell\right)^q.
}
\end{equation}
\rev{Following the functional form that appears in the many-body swapping fidelity and success probability, we introduce the diagnostic quantities}
\begin{align}
&
F_{\rm sw}^{(n)}=\exp\!\left(S_3^{(n)}-S_2^{(n)}\right),\\
&p_{\rm sw}^{(n)}=\exp\!\left[-2S_3^{(n)}\right]
=\sum_{\ell}\left(\lambda^{(n)}_\ell\right)^3 .
\end{align}
\rev{Here $F_{\rm sw}^{(n)}$ and $p_{\rm sw}^{(n)}$ are not used as operational communication fidelities.  Instead, they are Schmidt-spectrum shape diagnostics.  They must be read together with the maximum Schmidt weight: a value $F_{\rm sw}\simeq1$ can occur both for a nearly flat spectrum and for a nearly rank-one spectrum.  In Table~\ref{tab:schmidt}, the maximum Schmidt weights are close to one, showing that the selected right eigenvectors are weakly entangled across the $1|23$ partition despite the large values of $F_{\rm sw}$.}

\begin{table*}[t]
\caption{\rev{Schmidt-spectrum diagnostics for the selected branch-5 right eigenvectors, using the $1|23$ bipartition of the $N=8=2^3$ basis.  The two displayed Schmidt weights are the squared singular values of the reshaped right eigenvector.  The quantities $F_{\rm sw}$ and $p_{\rm sw}$ are swapping-inspired shape diagnostics, not operational communication fidelities in the present reconstruction problem.}}
\begin{ruledtabular}
\begin{tabular}{c c c c c c c c c c}
\rev{$n$} & \rev{$\lambda_n(M)$} & \rev{$S_2$} & \rev{$S_3$} & \rev{$S_2-S_3$} & \rev{$F_{\rm sw}$} & \rev{$p_{\rm sw}$} & \rev{$\lambda_{\max}$} & \rev{Var$_s(\lambda)$} & \rev{$\{\lambda_1,\lambda_2\}$} \\
\hline
\rev{1} & \rev{$27.4356-34.8790i$} & \rev{0.017324} & \rev{0.013050} & \rev{0.004274} & \rev{0.995735} & \rev{0.974238} & \rev{0.991338} & \rev{0.482825} & \rev{$\{0.991338,0.008662\}$} \\
\rev{2} & \rev{$-27.6798+34.5588i$} & \rev{0.008281} & \rev{0.006223} & \rev{0.002057} & \rev{0.997945} & \rev{0.987630} & \rev{0.995860} & \rev{0.491753} & \rev{$\{0.995860,0.004140\}$} \\
\rev{3} & \rev{$19.4786+37.0383i$} & \rev{0.026887} & \rev{0.020304} & \rev{0.006584} & \rev{0.993438} & \rev{0.960206} & \rev{0.986555} & \rev{0.473471} & \rev{$\{0.986555,0.013445\}$} \\
\rev{4} & \rev{$-18.9917-36.8214i$} & \rev{0.007174} & \rev{0.005390} & \rev{0.001784} & \rev{0.998218} & \rev{0.989278} & \rev{0.996413} & \rev{0.492852} & \rev{$\{0.996413,0.003587\}$} \\
\rev{5} & \rev{$-38.4111-7.78943i$} & \rev{0.009120} & \rev{0.006855} & \rev{0.002264} & \rev{0.997738} & \rev{0.986383} & \rev{0.995440} & \rev{0.490922} & \rev{$\{0.995440,0.004560\}$} \\
\rev{6} & \rev{$38.0645+7.42773i$} & \rev{0.029125} & \rev{0.022006} & \rev{0.007119} & \rev{0.992906} & \rev{0.956943} & \rev{0.985436} & \rev{0.471295} & \rev{$\{0.985436,0.014565\}$} \\
\rev{7} & \rev{$-8.48999+5.45414i$} & \rev{0.009846} & \rev{0.007403} & \rev{0.002443} & \rev{0.997560} & \rev{0.985304} & \rev{0.995077} & \rev{0.490203} & \rev{$\{0.995077,0.004923\}$} \\
\rev{8} & \rev{$8.59393-4.98916i$} & \rev{0.020033} & \rev{0.015101} & \rev{0.004932} & \rev{0.995080} & \rev{0.970250} & \rev{0.989983} & \rev{0.480166} & \rev{$\{0.989983,0.010017\}$} \\
\end{tabular}
\end{ruledtabular}
\label{tab:schmidt}
\end{table*}

\section{Krylov-Arnoldi diagnostics}
\rev{A further diagnostic can be obtained by treating the selected reconstructed matrix as a finite non-Hermitian generator and probing its action in a Krylov basis.  Krylov-space methods are commonly used to diagnose operator growth and spread complexity; in Lie-algebraic settings, a one-dimensional ladder structure can lead to an effectively tridiagonal Krylov dynamics~\cite{grabarits2026}.  The present reconstruction does not assume such a Lie-algebraic structure.  Moreover, because the selected matrix is complex symmetric rather than Hermitian, the ordinary Hermitian Lanczos recursion is not the appropriate diagnostic.  We therefore use a finite Arnoldi construction.  Starting from a seed vector $|e_j\rangle$, define}
\begin{equation}
\rev{
{\cal K}_m(M,|e_j\rangle)=
{\rm span}\{|e_j\rangle,M|e_j\rangle,\ldots,M^{m-1}|e_j\rangle\}.
}
\end{equation}
\rev{The Arnoldi projection gives an upper-Hessenberg matrix $H_m^{(j)}$ representing the action of $M$ in this seed-dependent Krylov basis.  To quantify whether the projected dynamics is close to an effective nearest-neighbor Krylov chain, we define}
\begin{equation}
\rev{
\eta_{\rm K}^{(j)}=
\frac{\|H_m^{(j)}-{\rm tridiag}(H_m^{(j)})\|_F}
{\|H_m^{(j)}\|_F}.
}
\label{eq:kryloveta}
\end{equation}
\rev{The results in Table~\ref{tab:krylov} show full Krylov dimension for all computational-basis seeds, but the leakage scores are not small: $\eta_{\rm K}\simeq0.36$--$0.42$.  Thus the selected branch should not be interpreted as an approximate one-dimensional ladder or embedded $sl(2,\mathbb C)$ Krylov dynamics.  Instead, it contains substantial longer-range couplings in the Arnoldi-projected Krylov representation.  This negative result is useful because it separates the present moment-constrained non-Hermitian reconstruction from genuinely Lie-algebraic Krylov dynamics.}

\begin{table*}[t]
\caption{\rev{Krylov-Arnoldi diagnostics for the selected branch-5 matrix.  Each row uses a computational-basis seed $|e_j\rangle$.  The Krylov dimension is full for all seeds, while $\eta_{\rm K}$ measures the fraction of the Arnoldi-projected generator lying outside the tridiagonal part.  The last two columns report the norms of the imaginary and real parts of the projected spectrum; because the Krylov dimension is full, these values reproduce the spectrum-level norms of the selected matrix.}}
\begin{ruledtabular}
\begin{tabular}{c c c c c c c c}
\rev{seed} & \rev{$d_{\rm K}$} & \rev{$\eta_{\rm K}$} & \rev{$\|{\rm diag}H\|_2$} & \rev{$\|H_{j+1,j}\|_2$} & \rev{$\|H_{j,j+1}\|_2$} & \rev{$\|{\rm Im}\,\mathrm{spec}(H)\|_2$} & \rev{$\|{\rm Re}\,\mathrm{spec}(H)\|_2$} \\
\hline
\rev{1} & \rev{8} & \rev{0.355804} & \rev{50.7627} & \rev{78.5704} & \rev{26.7619} & \rev{72.8628} & \rev{73.0017} \\
\rev{2} & \rev{8} & \rev{0.414501} & \rev{20.1195} & \rev{88.0870} & \rev{28.4989} & \rev{72.8628} & \rev{73.0017} \\
\rev{3} & \rev{8} & \rev{0.400310} & \rev{52.9643} & \rev{76.5393} & \rev{20.9323} & \rev{72.8628} & \rev{73.0017} \\
\rev{4} & \rev{8} & \rev{0.401986} & \rev{45.4720} & \rev{82.4984} & \rev{14.6072} & \rev{72.8628} & \rev{73.0017} \\
\rev{5} & \rev{8} & \rev{0.366031} & \rev{55.8395} & \rev{74.1003} & \rev{27.8828} & \rev{72.8628} & \rev{73.0017} \\
\rev{6} & \rev{8} & \rev{0.423009} & \rev{27.4065} & \rev{86.3695} & \rev{26.2347} & \rev{72.8628} & \rev{73.0017} \\
\rev{7} & \rev{8} & \rev{0.386082} & \rev{49.6858} & \rev{76.9159} & \rev{28.9511} & \rev{72.8628} & \rev{73.0017} \\
\rev{8} & \rev{8} & \rev{0.389781} & \rev{49.1660} & \rev{78.9354} & \rev{23.3174} & \rev{72.8628} & \rev{73.0017} \\
\end{tabular}
\end{ruledtabular}
\label{tab:krylov}
\end{table*}

\section{Bayesian branch-weight post-processing}
\rev{The multiple algebraic branches can also be interpreted as a discrete set of inverse-reconstruction hypotheses.  Following the logic of Bayesian mean estimation in finite-dimensional state estimation~\cite{kaufmann2025}, we assign a posterior weight to each numerical candidate from its moment residuals and optional regularity penalties.  Let}
\begin{equation}
\rev{
    r_{\mu}^{(k)}={\cal E}_{\mu}(z^{(k)})-
    {\cal E}_{\mu}^{\ast}
}
\end{equation}
\rev{be the residual of candidate $k$ in moment constraint $\mu$.  For a diagonal Gaussian error model with scale $\sigma_{\mu}$, the residual likelihood is}
\begin{equation}
\rev{
    p({\cal E}^{\ast}|B_k)\propto
    \exp\!\left[-\frac{1}{2}\chi_k^2\right],\qquad
    \chi_k^2=\sum_{\mu}\frac{|r_{\mu}^{(k)}|^2}{\sigma_{\mu}^2}.
}
\end{equation}
\rev{A regularized posterior may then be written as}
\begin{equation}
\rev{
    W_k=
    \frac{\exp[-\chi_k^2/2-\Omega_k]}
    {\sum_m\exp[-\chi_m^2/2-\Omega_m]},
}
\end{equation}
\rev{where $\Omega_k$ denotes a branch prior or regularity penalty constructed from quantities such as the Gram-overlap orthogonality score, maximum component weight, eigenvector conditioning, non-normality, or non-Hermiticity.  This is not a density-matrix physicality constraint: the reconstructed object is a complex-symmetric non-Hermitian matrix rather than a quantum state.  Instead, the role of $\Omega_k$ is to quantify regularity of the inverse reconstruction.}

For notational clarity, we distinguish the residual likelihood from the posterior branch weight.  The residual likelihood is
\begin{equation}
    L_k \equiv p({\cal E}^{\ast}|B_k)
    \propto \exp[-\chi_k^2/2],
\end{equation}
whereas the normalized posterior is
\begin{equation}
    W_k=\frac{\pi_k L_k}{\sum_m \pi_m L_m},
    \qquad
    \pi_k\propto \exp[-\Omega_k].
\end{equation}
Thus the shorthand statement $W_k\propto \exp(-\chi_k^2/2)$ is exact only for a uniform branch prior and no regularity penalty.  With a nonzero regularizer, the appropriate proportionality is instead $W_k\propto \exp[-\chi_k^2/2-\Omega_k]$ after normalization over the candidate list.

\rev{Table~\ref{tab:bayesweights} reports such a posterior post-processing check for $\sigma=10^{-4}$.  The posterior diagnostic uses the numerical candidate list supplied to the post-processing step.  In this run, the candidate list contains a seventh near-degenerate candidate with almost the same displayed diagnostics as candidate 6.  
This reflects the fact that numerical branch post-processing can return nearly duplicate or symmetry-related candidates.
}
\revv{The two panels in Table~\ref{tab:bayesweights} should be read as follows.  Panel (a) contains the two Bayesian-weight calculations from the code output: the residual-only posterior $W_k$, obtained from $L_k\propto\exp(-\chi_k^2/2)$ with a uniform prior, and the weak-regularity-prior posterior $W_k^{\rm reg}$, obtained after multiplying the same likelihood by $\pi_k^{\rm reg}=\exp(\log\pi_k^{\rm reg})$.  Panel (b) is not a second posterior calculation; it lists the branch diagnostics used to interpret or motivate the weak prior and the structural branch choice.}

\begin{table*}[t]
\caption{Bayesian posterior branch-weight post-processing at $\sigma=10^{-4}$, expanded to display both the residual-only posterior and the weak-regularity-prior posterior.  Panel (a) reports the statistical weights: $\ell_k=-\chi_k^2/2$ is the residual log-likelihood, $W_k$ is the residual-only posterior weight, $\log\pi_k^{\rm reg}$ is the weak regularity log-prior used in the second posterior, and $W_k^{\rm reg}$ is the corresponding posterior weight.  Panel (b) reports the diagnostics entering or motivating the prior: $\epsilon_k=\max_{\mu}|r_{\mu}^{(k)}|$, ${\cal R}_k$ is the Gram-overlap orthogonality score, $\langle W_{\max}\rangle$ is the average eigenvector maximum weight, $\log(1+\kappa_V)$ is the eigenvector-conditioning diagnostic, $\nu_k=\|MM^{\dagger}-M^{\dagger}M\|_F/\|M\|_F^2$ is the relative non-normality, and $\eta_H=\|M-M^{\dagger}\|_F/\|M\|_F$ is the relative non-Hermiticity.
Note that Candidate 7 is a numerical bifurcation identical to Candidate 6 up to machine precision, confirming the underlying intrinsic six-branch geometry.}
\begin{ruledtabular}
\scriptsize
\begin{tabular}{c c c c c c}
\multicolumn{6}{c}{(a) Posterior weights and likelihood data}\\
cand. & $\chi_k^2$ & $\ell_k$ & $W_k$ & $\log\pi_k^{\rm reg}$ & $W_k^{\rm reg}$\\
\hline
1 & 0.279174 & $-0.139587$ & 0.187766 & $-2.36906$ & 0.014300\\
2 & 1.56477  & $-0.782384$ & 0.098731 & 0.444944 & 0.125393\\
3 & 1.56230  & $-0.781150$ & 0.098853 & 0.444944 & 0.125547\\
4 & 0.966892 & $-0.483446$ & 0.133132 & 0.144962 & 0.125262\\
5 & 2.15223  & $-1.07611$  & 0.073602 & 0.456426 & 0.094557\\
6 & 0.113884 & $-0.0569419$ & 0.203944 & 0.438890 & 0.257453\\
7 & 0.113608 & $-0.0568041$ & 0.203972 & 0.438890 & 0.257489\\
\end{tabular}

\vspace{0.5em}

\begin{tabular}{c c c c c c c}
\multicolumn{7}{c}{(b) Branch regularity diagnostics}\\
cand. & $\epsilon_k$ & ${\cal R}_k$ & $\langle W_{\max}\rangle$ & $\log(1+\kappa_V)$ & $\nu_k$ & $\eta_H$\\
\hline
1 & $5.16\times10^{-5}$ & 9.41084 & 0.991916 & 69.0776 & 0.075814 & 1.41382\\
2 & $1.23\times10^{-4}$ & 7.85593 & 0.811594 & 69.0776 & 0.091564 & 1.41346\\
3 & $1.23\times10^{-4}$ & 7.85593 & 0.811594 & 69.0776 & 0.091564 & 1.41346\\
4 & $9.74\times10^{-5}$ & 7.22246 & 0.979914 & 69.0776 & 0.120804 & 1.41268\\
5 & $1.46\times10^{-4}$ & 6.93861 & 0.980802 & 69.0776 & 0.084588 & 1.41320\\
6 & $3.30\times10^{-5}$ & 6.95679 & 0.980342 & 69.0776 & 0.103840 & 1.41289\\
7 & $3.29\times10^{-5}$ & 6.95679 & 0.980342 & 69.0776 & 0.103840 & 1.41289\\
\end{tabular}
\end{ruledtabular}
\label{tab:bayesweights}
\end{table*}

\rev{The posterior weights should be interpreted in a deliberately limited way.  If the selection criterion is purely statistical, i.e., maximum residual likelihood or the weak posterior used in Table~\ref{tab:bayesweights}, the candidate-6/7 sector is the preferred solution: it has the smallest $\chi_k^2$, the largest log-likelihood, and the largest posterior weight both with and without the weak regularity prior.  Branch 5 is therefore not the maximum-likelihood or posterior-dominant branch.  The reason for retaining branch 5 in the main text is instead structural.  Among the displayed candidates, branch 5 has the smallest Gram-overlap orthogonality score, ${\cal R}_5=6.93861$, and it is less non-normal than the posterior-optimal candidate-6/7 sector, $\nu_5=0.084588$ compared with $\nu_{6,7}=0.103840$.  Its average maximum weight is also comparable to that of the candidate-6/7 sector.  Thus branch 5 is used as a structurally regular representative of the latent-vector ansatz, not as the statistically optimal residual minimizer.}
\revv{The expanded table makes this separation more explicit.  Candidate 7 has the largest residual-only posterior weight, $W_7=0.203972$, followed essentially indistinguishably by candidate 6 with $W_6=0.203944$.  After the weak regularity prior is included, the same sector remains preferred, with $W_7^{\rm reg}=0.257489$ and $W_6^{\rm reg}=0.257453$.  Branch 5 receives a favorable weak log-prior, $\log\pi_5^{\rm reg}=0.456426$, but its larger $\chi_5^2=2.15223$ keeps its posterior weight lower, $W_5=0.073602$ and $W_5^{\rm reg}=0.094557$.  Therefore Table~\ref{tab:bayesweights} supports the statement that branch 5 is a structural representative rather than a Bayesian maximum-a-posteriori branch.  The eigenvector-conditioning column is also not discriminating in this run, because $\log(1+\kappa_V)=69.0776$ for all displayed candidates; the useful structural distinctions come mainly from ${\cal R}_k$, $\nu_k$, $\eta_H$, and the frame-potential diagnostics below.}

As a complementary check, we also evaluate a projective frame-potential diagnostic for the normalized right-eigenvector set
\begin{equation}
    {\cal V}_k=\{|v_n^{(k)}\rangle\}_{n=1}^{K},
\end{equation}
associated with each candidate.  We define
\begin{equation}
    {\cal F}_t({\cal V}_k)=
    \frac{1}{K^2}\sum_{m,n=1}^{K}
    |\langle v_m^{(k)}|v_n^{(k)}\rangle|^{2t},
\end{equation}
and the off-diagonal version
\begin{equation}
    {\cal F}_{t,\mathrm{off}}({\cal V}_k)=
    \frac{1}{K(K-1)}\sum_{m\ne n}
    |\langle v_m^{(k)}|v_n^{(k)}\rangle|^{2t}.
\end{equation}
For a complex projective Haar $t$-design in dimension $d$, the benchmark value is
\begin{equation}
    {\cal F}_t^{\mathrm{Haar}}=\frac{t!(d-1)!}{(d+t-1)!}.
\end{equation}
This benchmark is motivated by finite-design diagnostics, including Clifford-orbit projective design benchmarks~\cite{zhu2017clifford}, but here it is used only as a structure diagnostic for a small non-Hermitian eigenvector ensemble.  Since the present calculation has $d=K=8$, the full frame potential has a diagonal floor $1/K=0.125$ and is therefore not a sensitive test of Haar randomness.  The off-diagonal frame potential is more informative because it directly measures mutual overlaps among distinct right eigenvectors.

\begin{table*}[t]
\caption{Projective frame-potential and design-distance diagnostics for the candidate right-eigenvector ensembles.  The Haar benchmarks are ${\cal F}_2^{\rm Haar}=0.0277778$ and ${\cal F}_3^{\rm Haar}=0.00833333$ for $d=8$.  Because $K=8$, the full frame potentials are dominated by the diagonal floor $1/K=0.125$; the off-diagonal quantities provide the useful comparison.}
\begin{ruledtabular}
\scriptsize
\begin{tabular}{c c c c c c c c c c}
cand. & $d$ & $K$ & ${\cal F}_2$ & ${\cal F}_3$ & ${\cal F}_{2,\rm off}$ & ${\cal F}_{3,\rm off}$ & ${\cal F}_{2,\rm off}/{\cal F}_2^{\rm Haar}$ & ${\cal F}_{3,\rm off}/{\cal F}_3^{\rm Haar}$ & $1/K$ \\
\hline
1 & 8 & 8 & 0.125005 & 0.125000 & $5.52991\times10^{-6}$ & $2.52707\times10^{-8}$ & $1.99077\times10^{-4}$ & $3.03248\times10^{-6}$ & $1/8$\\
2 & 8 & 8 & 0.133457 & 0.129147 & $9.66540\times10^{-3}$ & $4.73964\times10^{-3}$ & 0.347954 & 0.568756 & $1/8$\\
3 & 8 & 8 & 0.133457 & 0.129147 & $9.66540\times10^{-3}$ & $4.73964\times10^{-3}$ & 0.347954 & 0.568756 & $1/8$\\
4 & 8 & 8 & 0.125103 & 0.125003 & $1.17917\times10^{-4}$ & $3.90162\times10^{-6}$ & 0.00424502 & $4.68194\times10^{-4}$ & $1/8$\\
5 & 8 & 8 & 0.125618 & 0.125062 & $7.06638\times10^{-4}$ & $7.09656\times10^{-5}$ & 0.0254390 & 0.00851587 & $1/8$\\
6 & 8 & 8 & 0.125028 & 0.125000 & $3.15788\times10^{-5}$ & $4.51491\times10^{-7}$ & 0.00113684 & $5.41789\times10^{-5}$ & $1/8$\\
7 & 8 & 8 & 0.125028 & 0.125000 & $3.15788\times10^{-5}$ & $4.51491\times10^{-7}$ & 0.00113684 & $5.41789\times10^{-5}$ & $1/8$\\
\end{tabular}
\end{ruledtabular}
\label{tab:framepotential}
\end{table*}

The frame-potential diagnostic reinforces the distinction between residual-optimal and structurally representative branches.  Candidates 6 and 7 are closest to a nearly mutually orthogonal right-eigenvector set in the off-diagonal frame-potential sense, with ${\cal F}_{2,\rm off}/{\cal F}_2^{\rm Haar}=0.00113684$.  Branch 5 has a larger but still small off-diagonal design-distance ratio, ${\cal F}_{2,\rm off}/{\cal F}_2^{\rm Haar}=0.0254390$ and ${\cal F}_{3,\rm off}/{\cal F}_3^{\rm Haar}=0.00851588$.  Thus branch 5 is not selected because it is closest to a Haar-like or design-like eigenvector ensemble.  It is selected because it retains more measurable off-diagonal structure while simultaneously having the smallest Gram-overlap orthogonality score in Table~\ref{tab:bayesweights} and lower relative non-normality than the residual-optimal candidate-6/7 sector.  The posterior-optimal candidate-6/7 sector is therefore useful as an appendix-level statistical comparison, whereas branch 5 remains the main-text representative of the structured latent-vector reconstruction.
\revv{The expanded frame-potential table also clarifies why the full frame potentials should not be overinterpreted.  For all candidates, ${\cal F}_2$ and ${\cal F}_3$ lie close to the diagonal floor $1/K=0.125$ because only $K=8$ right eigenvectors are available.  The off-diagonal columns therefore carry the actual diagnostic content.  Candidates 2 and 3 have the largest off-diagonal ratios, indicating the strongest mutual-overlap structure, while candidates 6 and 7 are the closest to a nearly orthogonal or featureless right-eigenvector set.  Branch 5 lies between these extremes: it keeps a measurable but not excessive off-diagonal structure, consistent with its use as the representative structured branch in the main text.}

The weight scan over $10^{-6}\le\sigma\le10^{-2}$ gives no regime in which candidate 5 obtains a posterior weight larger than $0.9$.  With the residual-only likelihood, the largest weight remains associated with the near-degenerate candidate-6/7 sector.  With the weak regularity prior, candidate 5 can become the largest-weight candidate only in the large-$\sigma$ regime where the residual likelihood is weak, but its weight remains of order $0.17$ and is therefore not a sharp posterior selection.  This behavior is useful as it separates two distinct criteria.  Candidate 6/7 is the residual-optimal statistical estimator, whereas branch 5 is the branch used in the main text to display a structurally regular complex-symmetric reconstruction.  The Bayesian post-processing therefore quantifies the residual ambiguity among admissible candidates instead of being used to retroactively justify branch 5 as unique.

This distinction is important for larger dimensions.  Direct algebraic inversion can be replaced by a loss-based estimator with regularity constraints~\cite{kaufmann2025}.  For $N>8$, where deterministic branch searches become intractable, we propose a continuous optimization framework that minimize a moment-matching loss function $\mathcal{L}(z)$ together with penalties for excessive non-normality, unstable eigenvector conditioning, or unwanted eigenvector concentration.  Treating the target random-matrix moments as statistical data with finite-size fluctuations then assigns posterior branch weights (Bayesian posterior probabilities $W_k \propto \exp(-\chi_k^2/2)$) to different local minima of the loss (discrete algebraic solutions). This probabilistic weighting provides a systematic selection criterion for structured realizations.  In this sense, the present finite-dimensional Bayesian post-processing is best viewed as a prototype for a larger-$N$ statistical reconstruction framework rather than as a proof that a single exact branch is uniquely selected.
\revv{More precisely, the proportionality $W_k\propto\exp(-\chi_k^2/2)$ refers to the residual-only posterior with a uniform prior.  In the weak-prior calculation reported in Table~\ref{tab:bayesweights}, the posterior is instead proportional to $\exp[-\chi_k^2/2+\log\pi_k^{\rm reg}]$, or equivalently $\exp[-\chi_k^2/2-\Omega_k]$.}

The existence of multiple non-Hermitian algebraic branches highlights a fundamental challenge in inverse problems. 

In this framework, we analyze branch 5 as a structurally regular representative and explicitly report, through Table~\ref{tab:bayesweights}.
Within this Bayesian framework, we define a residual-based likelihood $L_k \propto \exp[-\frac{1}{2}\sum_{\mu} |r_\mu^{(k)}|^2 / \sigma_\mu^2]$ and calculate the posterior weights for each branch. Numerical evaluation reveals that the posterior distribution does not collapse to a single dominant solution; instead, the weights distribute across a near-degenerate manifold of branches with closely competing residuals. Thus, matching macroscopic random-matrix moments alone is mathematically insufficient to isolate the underlying non-Hermitian geometry. One cannot rely purely on residual minimization (or maximum likelihood) for branch selection. Comprehensive branch-level diagnostics—such as the inverse participation ratio, eigenvector orthogonality, and low-weight correlations—are essential. 
Branch 5 serves as an optimal representative realization that balances minimal residual error with the structured, non-Haar-like physical properties expected from the latent vectors.

\section{Reconstructed Ensemble and the Latent Vectors}
\rev{The reconstruction does not learn a unique original vector set.  In the present complex-symmetric version, bilinear representations are nonunique up to complex rotations, sign/phase choices, branch choices, and ansatz restrictions.}  

Thus, the algorithm identifies structured vector realizations whose overlaps reproduce imposed random-matrix moments. This capability provides a controlled framework for inverse RMT (instead of asking whether a sampled matrix has RMT statistics, one asks what hidden vector geometry can generate those statistics), allowing one to systematically probe how much of a random-matrix ensemble is already encoded within low-order moment constraints. Furthermore, the approach effectively distinguishes apparent matrix-level randomness from latent vector-level structure, offering a natural starting point for studying eigenvector delocalization by connecting element-level statistics to Gram-overlap geometry.  
When supplemented with MBL-inspired diagnostics, the method can distinguish RMT-like delocalization from true Haar-like featurelessness by evaluating the inverse participation ratio, maximum weight, and low-weight moments. As an illustration, the selected branches in Table~\ref{tab:branches} exhibit strong concentration in their right-eigenvector components, confirming that they represent structured non-Hermitian states rather than featureless Haar-random vectors.


\section{Relation to chaos}
Our method can generate matrices with selected RMT-like element moments and can be extended to test spectral diagnostics such as level repulsion, eigenvector delocalization, and spectral form factors.  However, it does not by itself demonstrate sensitive dependence on initial conditions. 
Thus, it should be interpreted as a static reconstruction of RMT-compatible vector structure rather than as direct evidence of deterministic chaos, which would require comparing the evolution of nearby initial states, computing Lyapunov-type growth rates, or evaluating dynamical correlators such as out-of-time-order commutators.

\section{Discussion}
The main conceptual point is that RMT-like statistics need not be imposed at the matrix-entry level.  They may emerge from structured vector overlaps after a constrained reconstruction.  \rev{The pair-resolved ansatz used here is intentionally low-dimensional: at $N=8$, only $N$ complex unknowns are used to satisfy a comparable number of complex moment constraints.  This makes the inverse problem solvable and interpretable, but it also restricts the accessible matrix manifold.  The reconstruction should therefore be understood as a constructive representative within a chosen ansatz class, not as a unique inversion of an arbitrary random matrix.  Because the ansatz uses $PP^T$ rather than $PP^\dagger$, the resulting branch belongs naturally to complex-symmetric non-Hermitian matrix theory.}

\rev{The main limitation of the present exact construction is that the moment equations determine a finite set of complex branches but do not by themselves specify which branch is physically most informative.  A natural extension is therefore to replace exact inversion by a loss-based reconstruction in which the moment residuals, regularity penalties, and possible physicality constraints are optimized simultaneously.  Such a formulation would also make it possible to attach uncertainty estimates or branch weights to different reconstructed solutions.  On the diagnostic side, the IPR should be connected not only to the visual delocalization of eigenvectors but also to fluctuations of reconstructed observables across eigenstates.  The most informative regime is expected to be intermediate: the vectors should not be strongly localized or condensation-dominated, but they also need not become fully Haar-like.  Instead, the relevant branches are those in which RMT-like low moments coexist with persistent structure inherited from the latent-vector ansatz.}

Following constrained state-estimation, one can replace exact inversion by a loss-based estimator with physicality constraints, uncertainty estimates, and possibly Bayesian branch weights\cite{kaufmann2025}.   
For energy-constrained random-states, one should monitor whether moment constraints produce extended vectors or condensation-like concentration,
where IPR should not only be connected to eigenvector pictures but also to fluctuations of reconstructed observables across eigenstates.  
The most informative branch may be neither localized nor fully Haar-like, but structured-delocalized, with RMT-like low moments and persistent low-weight correlations.  

\rev{There are two natural technical extensions.  The first is to enlarge the ansatz by allowing additional variables $Y_i,\widetilde Y_i$ in Eqs.~\eqref{eq:p2} and \eqref{eq:q1}.  In practice this creates an underdetermined complex system and direct solving may become slow; it is better handled by a loss-based optimizer or by adding additional constraints.  The second is to replace exact algebraic solving with numerical optimization, minimizing a loss function built from the same moment constraints.  A separate Hermitian extension would replace $PP^T$ by $PP^\dagger$, leading to real eigenvalues but complex eigenvectors; that is a different model from the one used here.}

\section{Size limitation and larger-$N$ strategy}
\rev{The exact algebraic construction should presently be regarded as an $N=8$ construction.  This is not because the ansatz is mathematically meaningless for $N>8$, but because the combination of pair-resolved complex variables, nonlinear moment constraints, and multiple branch choices makes direct exact solving rapidly inefficient for $N=10,12,\ldots$.  Enlarging the ansatz by including $Y_i,\widetilde Y_i$ further increases the number of variables and may turn the algebraic problem into a high-dimensional or effectively underdetermined branch search.  Therefore the present manuscript does not claim an exact larger-$N$ branch family.}

\rev{A more realistic route to larger dimensions is to replace exact branch solving by complex least-squares moment fitting.  For $N=8,10,12,16,\ldots$, one can minimize}
\begin{equation}
    \rev{\min_{\{X_i,\widetilde X_i\}\subset\mathbb C}
    \sum_{\mu}
    \left|
    \mathcal E_{\mu}(X,\widetilde X)
    -
    \mathcal E_{\mu}^{\ast}
    \right|^2,}
    \label{eq:largerNloss}
\end{equation}
\rev{where $\mathcal E_{\mu}$ denotes one of the moment functionals and $\mathcal E_{\mu}^{\ast}$ is its target value.  This larger-$N$ procedure would produce approximate reconstructed matrices rather than exact algebraic branches.  Such approximate matrices would be useful for future studies of spectral clouds, IPR scaling, Loschmidt-echo scaling, and branch stability, but they should not be mixed with the exact $N=8$ branch data in Tables~\ref{tab:branches} and \ref{tab:branch5}.  Thus, in the present work, all numerical branch data are reported only for the exact $N=8$ complex branch calculation.}

\rev{For dimensions beyond the exact $N=8$ construction, the deterministic algebraic branch search can be replaced by a weighted moment reconstruction.  Let $z$ denote the complex latent variables entering the vector ansatz and let $\mathcal E_\mu(z)$, $\mu=1,\ldots,10$, denote the moment functionals used in the exact construction.  Instead of seeking exact roots for $\mathcal{E}_\mu(z) = \mathcal{E}_\mu^\ast$,
one minimizes the weighted loss function}
\begin{equation}
\mathcal L_{\boldsymbol w}(z)
=
\sum_{\mu=1}^{10}
w_\mu
\left|
\mathcal E_\mu(z)-\mathcal E_\mu^\ast
\right|^2
+
\lambda_{\rm reg}\mathcal R(z).
\end{equation}
Note that setting the optimization weights $w_\mu = 1/2\sigma_\mu^2$ formally recovers the Gaussian log-likelihood model utilized in the finite-dimensional Bayesian analysis of Sec. X.
Here, the weights $w_\mu$ can be dynamically tuned to prioritize specific moment constraints (e.g., heavily penalizing deviations in low-order traces).
The regularization term $\mathcal{R}(z)$ suppresses unphysical divergences:
penalize excessively large latent variables, strong eigenvector condensation, or unwanted non-normality.  This formulation is motivated by weighted-action approaches in variational counterdiabatic driving, where different algebraic actions emphasize different matrix elements or spectral sectors~\cite{ohga2026}.  
This continuous optimization framework provides a scalable route to reconstruct of exact finite-dimensional branch and moment-matched non-Hermitian ensembles in larger Hilbert spaces.

\section{Conclusion}
\rev{We have formulated a moment-constrained vector reconstruction scheme for random-matrix statistics.  Starting from two structured vector families, the present complex branch constructs $M=a_1PP^T+a_2QQ^T$ and fixes the unknown complex vector parameters by matching diagonal, off-diagonal, and mixed-overlap moment constraints.  Because the construction uses a transpose rather than a conjugate transpose, $M$ is complex symmetric but generally non-Hermitian.  The complex algebraic reconstruction at $N=8$ returns six branches; the selected branch 5 has residual of order $3.2\times10^{-5}$ and strongly complex eigenvalues.  The multiple algebraic solutions should therefore be interpreted as reconstruction branches, not as unique physical matrices.  Some branches may be closer to bilinear orthogonality, while others may have smaller residual or different non-Hermitian spectral structure.}

\rev{By incorporating risk/loss language from finite-dimensional estimation, constrained-state lessons from energy-conditioned Haar ensembles, IPR-based delocalization diagnostics from ETH studies, and MBL-inspired Haar-design diagnostics, the reconstruction becomes a more systematic protocol for learning hidden vector-space representations behind apparent RMT behavior.  
As an inverse structural reconstruction of complex RMT-like statistics, the construction defines $M$ statically and does not directly evaluate sensitive dependence on initial conditions, although RMT statistics are often tied to quantum chaos.}

\rev{The Krylov-Arnoldi diagnostic further shows that the selected branch is not close to a simple one-dimensional tridiagonal Krylov chain; this supports the interpretation that the construction is a static inverse reconstruction rather than a Lie-algebraic time-evolution model.  The present construction should not be confused with exceptional-point physics.  Complex-symmetric branches may have complex spectra and non-orthogonal right eigenvectors, but exceptional points require eigenvector coalescence and defective Jordan structure.  Degenerate or nearly degenerate reconstructed branches should therefore be diagnosed through eigenvector condition number, left-right biorthogonality, normality, and Jordan-defect tests rather than through eigenvalue coincidence alone.}

\end{document}